# Toward reducing impact induced brain injury: Lessons from a computational study of army and football helmet pads[*]

William C. Moss[1], Michael J. King[1], and Eric G. Blackman[2]

[1]Lawrence Livermore National Laboratory, Livermore, CA 94551
[2]Department of Physics and Astronomy, University of Rochester, Rochester NY 14627

## Abstract

We use computational simulations to compare the impact response of different football and U.S. Army helmet pad materials. We conduct experiments to characterize the material response of different helmet pads. We simulate experimental helmet impact tests performed by the U.S. Army to validate our methods. We then simulate a cylindrical impactor striking different pads. The acceleration history of the impactor is used to calculate the Head Injury Criterion for each pad. We conduct sensitivity studies exploring the effects of pad composition, geometry, and material stiffness. We find that: (1) The football pad materials do not outperform the currently used military pad material in militarily-relevant impact scenarios; (2) Optimal material properties for a pad depend on impact energy; and (3) Thicker pads perform better at all velocities. Our analysis suggests that by using larger helmet shells with correspondingly thicker pads, impact-induced traumatic brain injury may be significantly reduced.

---

[*] *NOTICE: This article has been authored by Lawrence Livermore National Security, LLC ("LLNS") under Contract No. DE-AC52-07NA27344 with the U.S. Department of Energy for the operation of Lawrence Livermore National Laboratory ("LLNL"). Accordingly, the United States Government retains and the publisher, by accepting the article for publication, acknowledges that the United States Government retains a nonexclusive, paid-up, irrevocable, world-wide license to publish or reproduce the published form of this article, or allow others to do so for United States Government purposes.*

# 1 Introduction

About 80% of the 1.6 to 3.8 million annual U. S. civilian cases of head impact-induced traumatic brain injury (TBI), ranging from mild concussions to life threatening injuries, occur in motor vehicle accidents. About 20% occur in contact sports, primarily football [1]. Among deployed U. S. soldiers, IED attacks and other hazards have led to approximately 179,000 cases of TBI between 2000 and 2010 [2], caused by some combination of explosive blast exposure and impacts. Despite a better understanding of impact-induced TBI compared to blast induced TBI, helmet protection against impact-induced TBI is still deficient. Both combat helmets and sports helmets, particularly in youth sports, have lacked stringent impact TBI protection standards. The latter circumstance has led to the 2011 Children's Sports Athletic Equipment Safety Act, requiring improved regulation of youth helmets [3]. Although use of modern helmets has reduced concussions in the NFL, there is a dearth in understanding of the comparative efficacies of different helmet systems, thus indicating a need for such a study.

Here we present key results from a detailed computational study [4] comparing the impact response of selected football pad materials to U.S. Army helmet pads. This effort builds on our prior computational experience of comparing of the response of helmets and skulls to blasts versus impacts [5]. Although rotational accelerations can produce traumatic brain injury, we restrict our study to only linear accelerations from impacts because helmet pad materials and systems are limited in their ability to reduce injuries from non-impact rotational accelerations. Our aims are to elucidate the basic governing principles of pad protection against impact-induced TBI, compare the efficacy of football helmet pads with those currently used in U.S. Army helmets, and identify practical improvements that could be made to better protect soldiers or athletes. Our work is based on studying pad systems from four different manufacturers: Team Wendy, which produces the currently used U.S. Army pad; Oregon Aero, which produces the formerly used U.S. Army Pad; and Riddell and Xenith, which are both football helmet manufacturers.

We employed the following methods for our study: First, we conducted experimental compression tests at different compression rates of the various pad materials and of full pad systems in order to determine appropriate material properties for our computational models. We then conducted two sets of simulations. The first set of simulations was intended to validate our measured material properties and our simulation methodology. These simulations recreated a series of crown impact tests [6-8] conducted by the U.S. Army Aeromedical Research Lab [USAARL] on Army helmets with both Team Wendy and Oregon Aero pads. Good agreement between the simulation predictions and experimentally measured data validated our simulation methodology. Using this validated methodology, we designed a second set of simulations to isolate only the effects of the pad material on the impact response. These simulations consisted of a cylinder impacting a circular pad, removing the complicating effects of the helmet shell deformation, pad arrangement, impact angle, etc. These impact simulations allowed us to study the influence of pad material, composition, and geometry (thickness and area) on the impact response.

## 2   Experimental Characterization of Helmet Pads

Figure 1 shows the pads from four different manufacturers that were used in this study. Team Wendy makes the currently used U.S. Army pad, which consists of a bilayer foam (55.2 kg/m$^3$ and 54.6 kg/m$^3$) inside a waterproof polymeric wrapper and cloth bag. Oregon Aero makes the formerly used U.S. Army pad, which consists of a different bilayer foam (89.6 kg/m$^3$ and 86.9 kg/m$^3$) with a waterproof coating, also inside a cloth bag. Riddell manufactures the helmets most commonly used by the NFL. Riddell makes various types of helmets with different pads; the pads studied here were taken from Riddell's Revolution$^{TM}$ helmet and consist of a bilayer foam (97.7 kg/m$^3$ and 66.9 kg/m$^3$) inside an inflatable casing with air relief channels connecting the pads (referred to by Riddell as their Custom Fit$^{TM}$ Liner). Xenith also makes various different football helmets. The pads studied here were taken from Xenith's X1 helmet, and consist of an elastically buckling air-filled structure, with a thin foam cushion for comfort.

Material properties for the various pads were obtained from low and high strain rate unconfined compression tests performed at Lawrence Livermore National Lab [LLNL]. For the three pad systems that consist of a bilayer foam inside a covering (the Team Wendy, Oregon Aero, and Riddell pads), tests were performed on both the individual foam materials, and the complete pad system, with and without covering. Since the Xenith system achieves a foam-like behavior through the use of an engineered elastically buckling structure rather than a combination of different individual foam materials, only complete pad system tests were performed on the Xenith pads. The tests were used to determine material properties for numerically simulating the impact response of the pads. All tests were conducted at room temperature.

Our analysis code includes a non-linear rate-sensitive foam material model developed by Puso [9] to represent the pad materials. The properties of each foam in each pad were obtained by fitting this model to the experimental data. For the Xenith pads, the response of the complete pad system resembles that of a bilayer foam, so these pads were modelled as a bilayer foam with properties adjusted to fit the measured Xenith pad response. The data from the foam characterization experiments and further details regarding the material model are given in Appendix B of [4].

The air trapped by the waterproof coverings on the Team Wendy and Oregon Aero pads did affect the pad response. The effects of this trapped air were included in the simulations by placing a compressible material with appropriate properties in parallel with the foam material. More detail regarding this methodology is described in Appendix C of [4]. Although the Riddell system also has a pad covering, our compression tests indicated that its covering does not affect the pad response for the impact velocities considered (<6.10 m/s), whether pre-inflated or not. This is presumably due to the relief channels between adjacent pads that allow air to flow from one pad to its neighbours.

## 3  Validation simulations of USAARL experiments

### *3.1  Simulation Methods*

To establish the validity of our computational methods and foam material models, we first simulated crown impact experiments performed previously by the USAARL ([6-7]) and compared our simulations to the experimental data [8]. Figure 2 shows the typical experimental configuration [6] consisting of a magnesium alloy headform from Cadex, Inc. [10] secured by a tightly cinched chinstrap to a Kevlar™ Advanced Combat Helmet [ACH] shell containing Team Wendy or Oregon Aero foam pads. The headform is attached to a vertical track by a metal arm and is on a gimballed mount that allows selection of arbitrary impact points on the helmet. However, all experiments that we simulated were crown impact experiments. The headform is constrained by the arm and track to move only in the vertical direction. The weight of the headform and arm is 5 kg (approximately the mass of a human head). The helmet/arm is dropped onto a steel anvil from sufficient heights to overcome the track friction and achieve desired impact velocities ranging from 1.52-6.10 m/s, which is considered by the Army to be the range of militarily-relevant non-lethal impact velocities [6-7].

Data for two different impact velocities (3.05 m/s and 4.31 m/s) were available, for both the Team Wendy and Oregon Aero pads. For the Team Wendy experiments, complete time history data for the motion of the headform was available [8], allowing simulation predicted and experimentally measured velocity and acceleration traces to be compared directly. For the Oregon Aero experiments, only peak acceleration data were available [6].

The experiments were simulated in 3-D using the finite element analysis software PARADYN [11], which is a massively parallelized version of the Lawrence Livermore National Laboratory [LLNL] developed DYNA3D analysis software. This analysis code is specialized for the simulation of impacts and other dynamic events.

The simulation geometry is shown in Figure 3 and incorporates accurate geometric representations of the experiment components. An X-ray computed tomography (CT) scan of an actual ACH was used to develop the 3D geometry of the helmet shell in the model, which was then smoothed and meshed. The headform geometry was created from drawings available from the manufacturer [10]. The pad geometry was obtained by measuring manufacturer-supplied samples. Further details pertaining to the model geometry are given in Appendix A of [4]. Figure 3a shows a cutaway along the symmetry plane. An interior view of the complete helmet and pad mesh is shown in Figure 3b.

Material properties for the foam pads were obtained from experiments as described in the preceding section. Material properties for the Kevlar™ helmet shell were obtained from the literature [12]. The magnesium K1-A alloy headform was modelled as a linearly elastic material [13]. Room temperature conditions were assumed for all simulations.

In the simulation, the helmet, pads, and headform have initial downward velocities that are equal to the test velocities in the USAARL experiments, while the steel anvil is constrained to be motionless on its bottom face. The simulation begins approximately 0.5 ms prior to anvil contact. Because the pads do not initially conform to the curved surfaces of ACH shell and headform, there are initially small gaps at the interfaces. Early in the simulation, the impact loads displace the pads so that they conform to the headform and shell geometries. All contact in the simulation is modelled as sliding with friction (friction coefficient $\mu = 0$ at the headform-pad interface, and $\mu = 1$ at the pad-helmet and helmet-anvil interfaces).

### *3.2 Simulation Results*

Figure 4 shows comparisons of the simulated and experimentally measured headform velocity (measured at the headform center-of-mass) for 3.05 and 4.31 m/s impact velocities [7-8], with the helmet containing size #6 Team Wendy pads (which are 1.91 cm thick). Two sets of experimental data are shown in the figure to show the typical experimental scatter. There are only slight differences between the simulations and the experimental data. The agreement of the predicted and measured rebound velocities shows that the overall energy transfer and dissipation is modelled correctly.

Figure 5 shows the simulated and experimentally measured headform center-of-mass accelerations. All traces have been filtered at 1500 Hz to eliminate noise. The simulation results exhibit some oscillations in the acceleration history that are more severe than those in the experimental data. This is a relatively common occurrence in the simulation of impact events. These oscillations may be due to the loose fit of the internal components in the simulation model compared to the "tightly cinched" [14] chinstrap in the experimental setup, or other differences in the constraints on the headform in the experiment versus the simulation. Another possibility is that there may be damping mechanisms in the actual experiment that suppress higher frequency rattling of the headform, which were not modelled in the simulations.

These oscillations are not overly significant when evaluating the severity of the impact as it relates to injury. In general, head injury depends on both peak acceleration and duration of the acceleration event. The simulation predicts both of these metrics accurately despite the oscillations. In particular, the peak acceleration values (the metric reported by USAARL when evaluating a helmet pad system) predicted by the simulation lie within the experimental scatter of the measured values, as shown in Table 1.

More sophisticated metrics for evaluating impact severity exist. In particular, we employ the Head Injury Criterion (HIC) [15] in this study, which is a well established metric that quantifies the severity of the linear time-dependent acceleration history in an impact event. Refer to the Appendix for further detail regarding the HIC and our reasons for using it in this study. For the 4.31 m/s experiment, the HIC computed for the experiment and the simulation agree well. For the 3.05 m/s experiment, the oscillations in the simulation slightly reduce the average acceleration during the event, resulting in a lower HIC.

Data for peak acceleration in crown impacts at the same velocities, using Oregon Aero pads, were also available from USAARL [6-7]. We repeated the simulations replacing the Team Wendy pads with Oregon Aero pads, which are more compliant, and again found the simulation predictions to lie within the experimental scatter of the data, as shown in Table 1. No time history traces for the Oregon Aero impact experiments were available.

By showing that our simulations accurately predict experimentally measured velocities, peak acceleration, and acceleration duration in a helmet impact event, we have demonstrated that our material representations of the pad systems are reasonable and that our simulation methods are robust.

## 4 Pad response in cylinder impact simulations

### 4.1 Simulation Methods

Football helmet shells are larger and cover more of the head than an ACH, and employ pads that are nearly double the thickness of ACH pads. Football helmet shells deform and absorb energy differently than an ACH under impact. The distribution of pads within the helmets also differs, as shown in Figure 6. These differences can obscure the effects the pad materials have on head injury when full helmet simulations or experiments are conducted. To isolate the effects of the pad materials and to create a standardized comparison between them, the simplified geometry shown in Figure 7 was developed, instead of a more realistic head and helmet geometry. The simplified geometry consists of a cylindrical impactor hitting a circular pad that rests on a frictionless rigid surface. An initial velocity is imparted to the impactor and the velocity and acceleration history of the center of the impactor is tracked.

The Head Injury Criterion (HIC) associated with this acceleration history is computed and used to compare pad performance. The definition of the HIC and the reasons for using it as a comparative metric are discussed in the Appendix. Note that the HIC calculated for this simplified geometry is not used to predict the probability of injury associated with the use of a given pad, since the acceleration history and corresponding HIC associated with a geometrically accurate head and helmet system would be different. Factors such as deformation of the helmet shell, off-axis effects, sharing of load between multiple pads, effects of the chin strap, etc. could help mitigate or exacerbate the impact. Consequently, when we state that a given pad produces a certain HIC in the cylinder impact simulation, this does not mean that a helmet containing those pads would necessarily produce the same HIC.

The impactor has a mass comparable to a typical human head (5 kg) and is made of magnesium alloy like the experimental headforms used by USAARL. The impactor has a diameter 10% greater than the pad so that edge effects do not affect the simulation. Impact velocities range from 1.52 to 6.10 m/s, which is a range of impact velocities of interest for non-lethal military impact scenarios [6-7].

In order to isolate the pad material response from the effects of differently sized pads, the different pads are scaled to a thickness of 1.91 cm and a diameter of 12.7 cm (corresponding to a pad area of 126.68 $cm^2$) in most simulations. These dimensions were chosen for convenience and are consistent with a size 6 ACH crown pad. Crown pad dimensions were chosen because crown impacts are the simplest mode of impact in an ACH: during a crown impact, only the single crown pad with a well characterized area supports the head, while in front, side, or back impacts, multiple smaller pads at different angles support the head and the actual pad engaged is more complicated to determine. Because injuries are not limited to crown impacts but can occur in a variety of orientations, simulations that investigated the effects of changing the pad area and thickness were performed to determine the sensitivity of the pad response to pad geometry.

For bilayer foam pads simulated with a thickness different from an actual pad (e.g. the Riddell pads), the two simulated foams are used in the same proportion as in the actual system. For the Team Wendy and Oregon Aero pads, the effects of the trapped air were included by adding additional stiffness to the pad system, as described in Appendix C of [4]. Room temperature conditions were assumed, as data for the foam response at extreme temperatures were not available.

*4.2   Results*

*4.2.1   General response of the foam pads*

The simulations demonstrated the general response of the foam pads to impact. Foams have a characteristic response to compression, represented schematically in Figure 8, where force is plotted as a function of compression. The area under the curve gives the impact energy absorbed by the foam. After a brief elastic response, the cell walls of the foam begin to buckle and the foam crushes under constant load. When the foam densifies (i.e. compresses to the point where its inner cavities completely collapse), the force required to compress it further increases dramatically.

This general response demonstrates why a particular foam is suitable for protecting only against impacts of a given energy. A hard foam with a high crush plateau will dissipate all the impact energy without densifying. However, it does so by applying a larger force than is needed, and hence causes greater acceleration. A softer foam that is "just right" uses the entire crush plateau to dissipate the impact energy, and does so at a lower force (causing lower acceleration). If a foam is too soft for a given impact energy, it densifies (or "bottoms out") before this energy is absorbed by the crush and the peak force (and hence, peak acceleration) becomes large. For a given geometry, a single foam is therefore approximately optimal for only one specific impact energy, so the use of a multilayer foam can make pads more effective over a wider range of impact energies. We say "approximately" because the foam response is generally visco-elastic. As the impact velocity, and hence the energy, increases, the plateau stress also increases and accommodates some increased energy (although perhaps not optimally). If a soft foam is mixed with a harder foam, the softer foam absorbs energy before the harder foam deforms significantly. If the soft foam can absorb all the impact energy, the impact is mitigated with relatively low force and low acceleration. For impacts with greater energies, once the softer foam densifies, the remaining energy will be absorbed by the compaction of the harder foam. Ideally, this will result in a lower acceleration and force than if the entire pad were composed of just a soft foam that completely densified, or just a hard foam that applied too much force in response to the initial, high-speed impact.

*4.2.2 Comparison of the three foam pad systems*

Figure 9 compares the cylindrical impact simulation results for the three systems composed of bilayer foam pads (Team Wendy, Oregon Aero, and Riddell) over a range of velocities from 1.52 to 4.57 m/s. The x-axis is the relative kinetic energy of the impactor normalized to a 3.05 m/s impact. The y-axis is the HIC (see the Appendix) associated with the acceleration history of the impact. All simulations used the same pad geometry— that of a #6 ACH crown pad (1.91 cm thick and a 12.7 cm diameter, with a total area of 126.68 cm$^2$). The fractions of soft and hard foam in each simulation were the same as in the actual pad systems (50%-50% for the Team Wendy and Oregon Aero, and 30%-70% for the Riddell [16]).

The Riddell system is the hardest of the three pad systems, while the Oregon Aero is the softest. At low speeds (3.05 m/s or less), the Team Wendy and Oregon Aero pads produce an almost identical HIC, while the Riddell pad produces a slightly larger HIC. As the impact velocity is increased to 4.57 m/s, the Oregon Aero pad reaches a point when both component foams completely densify and the peak acceleration and HIC increase greatly relative to the other two pad systems. At 4.57 m/s the Team Wendy pad starts to densify, whereas the harder Riddell pad still resists the impact in the plateau region of its harder foam. As a result, the Team Wendy pad produces a similar HIC to the Riddell pad.

At 6.10 m/s (not shown in the figure), the Riddell pad system begins to densify, and all three pads have exceeded their effective impact absorption capabilities. The Oregon Aero pad produces the highest HIC in this case, followed by the Team Wendy pad and then the Riddell pad. However, impacts that compress the pads beyond densification would likely be lethal, and lie outside the scope of this study. Overall, for identical pad geometries, the Oregon Aero pads and the Riddell pads provide no apparent benefit over the Team Wendy pads currently used in the ACH, for non-lethal militarily relevant impact scenarios.

It should be noted that we do not claim that a 6.10 m/s impact is lethal to a person wearing an actual helmet. A real helmet system is probably more compliant than the simplified cylindrical geometry, so compressing the pads beyond densification in a real helmet system would likely require even higher impact speeds. While our results indicate that a single 1.91 cm thick crown pad alone is not likely to protect against injuries for impacts above 4.57 m/s, an entire helmet system that spreads the impact across multiple pads and includes the effect of the deformable helmet shell could be more effective. The cylinder impact simulations are useful for comparing one material to another but do not necessarily predict the likelihood of injury for distinct full helmet systems.

*4.2.3   Effects of pad thickness*

We next consider the importance of pad thickness. Thicker pads employ a greater "stopping distance" and hence can dissipate a given amount of energy with less force (and hence a lower peak acceleration, HIC, and lower chance of injury). Figure 10 compares the cylinder impact simulation results for Team Wendy pads scaled to different thicknesses ranging from 1.52 cm to 3.81 cm, over a range of velocities from 1.52 to 4.57 m/s. All simulations used a 12.7 cm pad diameter and an area of 126.68 cm$^2$. The fractions of soft and hard foam in each simulation were the same as in the actual #6 pad system (50%-50%). The thickness of the #6 ACH crown pad is 1.91 cm. At every impact velocity, a thicker pad produces a lower HIC, although the effect is most significant at larger velocities. For example, at 4.57 m/s, increasing the pad thickness by 0.38 cm from 1.91 cm to 2.29 cm reduces the HIC from 917 to 665, and increasing by another 0.38 cm to 2.67 cm reduces it to 528.

For a given impact energy, there is a limit to the benefit imparted by thickening the pads, evident as the curves coalesce at higher thicknesses. At 3.05 m/s, little benefit is attained with pads thicker than 1.91 cm. At 4.57 m/s, little benefit is attained with pads thicker than 3.05 cm. This is likely because the main benefit of a thicker pad is that it absorbs more energy before it densifies. When a pad is thick enough to avoid densification at a given energy, additional thickness reduces the HIC by only small amounts.

Overall, the dependence of the HIC on thickness suggests that wearing helmet shells that are at least one size larger and have correspondingly thicker pads could reduce injuries dramatically for severe impacts.

### 4.2.4  *Impact Response of the Xenith Air-Filled Pad*

For the Team Wendy, Oregon Aero, and Riddell pads, our methodology was to scale them all to a common pad thickness and diameter comparable to an ACH #6 crown pad. But unlike the other pads, the Xenith pad is not a bilayer foam. It is a complex air-filled structure designed to elastically buckle at a given load and expel air at a controlled rate. Each pad is approximately 5.08 cm in diameter and 3.68 cm thick and its impact response if it were thinner cannot be predicted by simply scaling down its thickness.

However, because the other three pad systems are bilayer foams, for which the material behavior has been characterized, they can be scaled up to the larger thickness of the Xenith pad. The response of the Xenith pad system can then be compared to the response of the other three pad systems at the same thickness.

An actual 5.08 cm diameter Xenith pad cannot be physically scaled up to the 12.7 cm diameter of an ACH crown pad. However, a simulation of a 12.7 cm diameter pad with Xenith-like behavior can be thought of as equivalent to a number of Xenith pads with a combined total area of 126.68 $cm^2$ acting side by side. Hence the response of all four systems can be compared at the same area.

Figure 11 compares the cylinder impact simulation results for the Team Wendy, Oregon Aero, and Riddell bilayer foam pad systems to the Xenith pad system for velocities from 1.52 to 6.10 m/s. All simulations used the same pad geometry: 3.68 cm thick and a 12.7 cm diameter (126.68 $cm^2$ area). The fractions of soft and hard foam in the three foam pad systems matched the actual pad systems (50%-50% for the Team Wendy and Oregon Aero, and 30%-70% for the Riddell [16]).

The results in Figures 9 and 11 are similar but the pad systems in Figure 11 are thicker, so the HIC values are lower at all velocities than for the same systems in Figure 9. The Team Wendy and Oregon Aero foams are the softest and have the lowest HIC values at low speeds. The soft Oregon Aero pads begin to densify at impacts of about 4.57 m/s, while the harder Team Wendy pads begin to densify at somewhat higher speeds. Note that because of the increased thickness, densification occurs at higher speeds for all the pads in Figure 11 relative to Figure 9. The Riddell pad is harder than the Team Wendy and Oregon Aero pads and produces a higher HIC at low speeds, but has still not densified at 6.10 m/s and produces the lowest HIC.

For the same 3.68 cm thickness and 126.68 cm² total pad area, the response of the Xenith pads is significantly stiffer (i.e. the pads are harder) than any of the foam pads. Consequently, they produce a higher HIC at every impact. However, they never are compacted to the point where their response stiffens (the equivalent of the foams densifying).

As was the case for the 1.91 cm thick pads in Figure 9, for identically sized pads, Figure 11 highlights that the Oregon Aero pads and the football pads (Riddell and Xenith) do not provide any apparent benefit over the Team Wendy ACH pads for the militarily relevant impact velocity ranges considered here, if the same pad geometries are used.

*4.2.5  Effect of Pad Area on Pad Impact Response*

Having compared pads of identical geometries, we note that different helmet systems may employ different numbers of differently sized pads in different arrangements. The amount of pad area that stops a given impact is not necessarily the same for the different systems, or even for differently oriented impacts in the same helmet system. For example, in a Riddell football helmet, the pads protecting against crown impacts cover a smaller area than typical ACH crown pads. It is therefore important to discuss the effect of pad area on impact response.

Figure 12 compares the cylinder impact simulation results for a Team Wendy crown pad with area 126.68 cm², Riddell pads with areas of 126.68 cm² and 63.34 cm², and pads with Xenith behavior and areas of 81.07 cm² and 40.54 cm² (corresponding to four and two Xenith pads, respectively). The reduced pad areas for the Riddell and Xenith pads roughly correspond to actual pad areas used in football helmets: the Riddell crown pad has approximately half the area of an ACH crown pad, while in the Xenith helmet the pads are positioned so that impacts from various angles are absorbed by between two and four pads. All pads had the same thickness of 3.68 cm. The fractions of soft and hard foam in the two foam pad systems were the same as in the actual pad systems (50%-50% for Team Wendy and 30%-70% for Riddell [16]).

Reducing a pad's area effectively makes it softer, but also causes it to densify sooner. When the Riddell pad's area is reduced by 50% (roughly comparable to the area that actually supports the crown of the head in a football helmet), its response at lower impact speeds is comparable to the Team Wendy pad system. However, the reduced area Riddell pad densifies above 4.57 m/s and its HIC value at 6.10 m/s is significantly higher than the Team Wendy pad or the full area Riddell pad.

The reduced area Xenith pads also produce a lower HIC at low speeds than the full area Xenith pad. Two 5.08 cm diameter Xenith pads produce a comparable HIC to the 12.7 cm diameter Team Wendy pad at 1.524 and 3.05 m/s. They produce a slightly higher HIC at 4.57 m/s, but don't bottom out in the velocity range in the figure, so that at 6.10 m/s they are comparable to the Team Wendy pad, which is densifying. The response of four 5.08 cm diameter Xenith pads lies between the two-pad response and the 126.68 cm$^2$ area Xenith response.

When pad areas are reduced, the football pad systems produce HIC values that are comparable to the Team Wendy system at low speeds, but only the reduced area (2 pad) Xenith system is comparable at high speeds. However, at no speed does performance of the football pads exceed that of the Team Wendy ACH system for these militarily relevant impact scenarios. This does not imply that the Team Wendy pad system would be better than the NFL systems in football-relevant impact scenarios, since football-relevant impacts may be different in character (helmet shell compliance, impact velocity, mass of the impacting objects, etc.) from militarily relevant impacts. For example, in combat, if a soldier is knocked to the ground, or against an object, or suffers a glancing blow to the head by debris or shrapnel, it may be appropriate to consider only the mass of the head when determining the impacting mass that the pads must absorb. However, in many football-relevant impact scenarios, a significant fraction of the entire body mass of the player may be driving the head into another object, and hence the impact energy for such hits that the pads need to absorb may be greater [18]. The thicker and harder NFL pads may therefore be more effective for higher energy impacts.

## 5    Discussion and Conclusion

We have used computational simulations to study how different helmet pad systems mitigate head impacts. We considered the Team Wendy foam pads currently used in the ACH, Oregon Aero foam pads formerly used in the ACH, Riddell foam football helmet pads, and Xenith football helmet pads.

To obtain material properties needed for the simulations, we performed experimental stress-strain measurements over the strain rates of interest to obtain the material response of the different pad foams and of the complete pad systems. We used CT scans of the helmet shell to develop accurate geometric computational models of the helmet. Combining the material models of the foams and geometric model of the helmet shell, we computationally recreated a series of crown impact experiments conducted at USAARL, and showed that the simulations capture the essential features of the experimental data.

Having validated our computational approach, we developed a simulation geometry for isolating the effects of the foam material from other effects, such as interactions between pads and deformations of the helmet shell. A simplified geometry was necessary to remove the influence of pad shape and helmet shell differences, and to standardize the comparison between pad materials. For this purpose, we simulated a cylinder with a mass comparable to a human head impacting against circular pads of the different materials, each comparable in size and geometry to an ACH crown pad. We conducted a range of sensitivity studies examining how the impact response depended on foam material, pad thickness, and pad area.

At lower impact velocities softer pads perform better, and at higher impact velocities harder pads perform better. Reducing the area of a pad is equivalent to making it softer. Thicker pads perform better at all velocities, but especially at high velocities. This suggests an immediate low-cost and practical mitigation strategy for lessening the severity of impact-induced traumatic brain injury for soldiers: by using helmet shells that are at least one size larger with correspondingly thicker pads, the injuries from impacts, especially severe impacts, may be reduced significantly. Our results also show that for comparable geometries and for room-temperature militarily relevant impact scenarios, neither the Riddell, the Xenith, nor the Oregon Aero pads outperform the Team Wendy pads currently used in military ACH systems.

More broadly, our work also exemplifies how numerical modelling, simulation, and analysis provide a versatile complement to experimental testing to improve head impact protection equipment and could yield overall cost savings during product design, development, and testing. A large number of simulations can be run quickly and cost effectively compared to experiments, and the simulations can illuminate the physical mechanisms and principles that guide conceptual strategies for improvement. This is exemplified in our discussion of Figure 8, which illustrates how an ideal foam should neither be too soft nor too stiff and how multi-layer foams are needed to ensure this optimal condition can be met at more than a single impact velocity.

Finally, we emphasize that our study focused on the performance of individual pads in standardized tests, which is distinct from studying the efficacy of different complete helmet systems. It is important to identify the principles of protection of each constituent of a helmet separately to know how changing each constituent will influence an overall helmet system.

We acknowledge funding from the U.S. Army and the Joint IED Defeat Organization [JIEDDO] through COL R. Todd Dombroski, DO, JIEDDO Surgeon. We thank LLNL staff members B. Cracchiola for loaning us his ACH helmet, D. Urabe for performing the compression tests on the manufacturer supplied pads and component foams, and W. Brown for CT scanning the ACH helmet and post-processing the files. We also thank N. Kraemer (Riddell), S. Reynolds (Xenith), R. Szalkowski (Team Wendy) and T. Erickson (Oregon Aero), for supplying pad and foam samples used in this study. We thank the U.S. Army Aeromedical Research Laboratory [USAARL] for providing us data from impact tests on infantry helmets. This work performed under the auspices of the U.S. Department of Energy by Lawrence Livermore National Laboratory under Contract DE-AC52-07NA27344.

**Appendix: Use of the Head Injury Criterion as a metric for measuring impact response**

Use of only the peak acceleration to evaluate the severity of an impact, or the effectiveness of a pad at protecting in that impact, is a flawed approach. Larger peak accelerations over very short durations are known to be less dangerous than lower accelerations over longer durations. The peak acceleration can also be very sensitive to experimental or computational variations which could produce misleading injury risks if, for example, high accelerations occur only for a very short time. A related issue is that acceleration data, whether obtained from an experiment or a computational simulation, is generally noisy and must be filtered, and the peak acceleration depends on the type of filter and the filtering frequency used. Two experiments could produce very similar acceleration histories with similar likelihoods of injury, but if the experimental data were filtered at different frequencies, dramatically different peak accelerations could be reported.

Instead, we evaluate impact severity using the Head Injury Criterion (HIC). This is a well-established metric developed from experimental data that quantifies the severity of linear acceleration upon impact, with respect to the likelihood of injury [15]. For a given linear acceleration history $a(t)$, the HIC is defined as:

$$HIC = \max_{t_1, t_2} \left\{ \left[ \frac{1}{t_1 - t_2} \int_{t_1}^{t_2} a(t) dt \right]^{2.5} (t_1 - t_2) \right\}, \tag{1}$$

where the acceleration $a(t)$ is in units of gravity (1 G = 9.8 m/s$^2$) and the initial time $t_1$ and the final time $t_2$ are in units of seconds. A bounding time interval $t_2 - t_1 \leq 0.015$ seconds is commonly used to ensure that the HIC is computed temporally close to the acceleration peak. Note that if large accelerations are very short in duration, the corresponding HIC can be less than for lower accelerations that last much longer.

The value of the HIC can be related to the probability of injury [17], provided that it is calculated from an acceleration trace representing a realistic impact event. A HIC of 1000 is considered to be life threatening and corresponds to an 18% chance of severe injury, a 55% chance of serious injury, and a 90% chance of moderate injury in the average adult. A HIC of 600 is considered to be the threshold for moderate injury and corresponds to an 18% chance of serious injury and a 50% chance of moderate injury. A HIC of about 300 is the threshold for minor concussions (approximately a 50% chance).

Because the HIC considers both peak acceleration and duration, it more accurately reflects the severity of an impact than peak acceleration alone. Because it integrates the acceleration history, it is far less sensitive to highly transient variations in acceleration, noise, and filtering method than peak acceleration. For these reasons, we employ the HIC to characterize the severity of an impact and to evaluate the relative effectiveness of various pad materials. However, we do not attempt to use the HIC to estimate probability injury for scenarios where the simulation geometry does not correspond to a realistic impact.

| Impact Velocity (m/s) | USAARL Data | | | | Simulation | | | |
|---|---|---|---|---|---|---|---|---|
| | Team Wendy | | Oregon Aero | | Team Wendy | | Oregon Aero | |
| | Peak G | HIC | Peak G | HIC | Peak G | HIC | Peak G | HIC |
| 3.048 | 86±2 | 213±6 | 77±5 | — | 83 | 156 | 77 | 126 |
| 4.310 | 123±8 | 457±16 | 150±24 | — | 126 | 441 | 131 | 437 |

**Table 1 – Peak acceleration and HIC values for USAARL experiments and corresponding simulations. Note that the data presented here omits one significant outlier for the Team Wendy impact at 3.05 m/s, where the peak G's were nearly twice as large as the measured values in three other tests.**

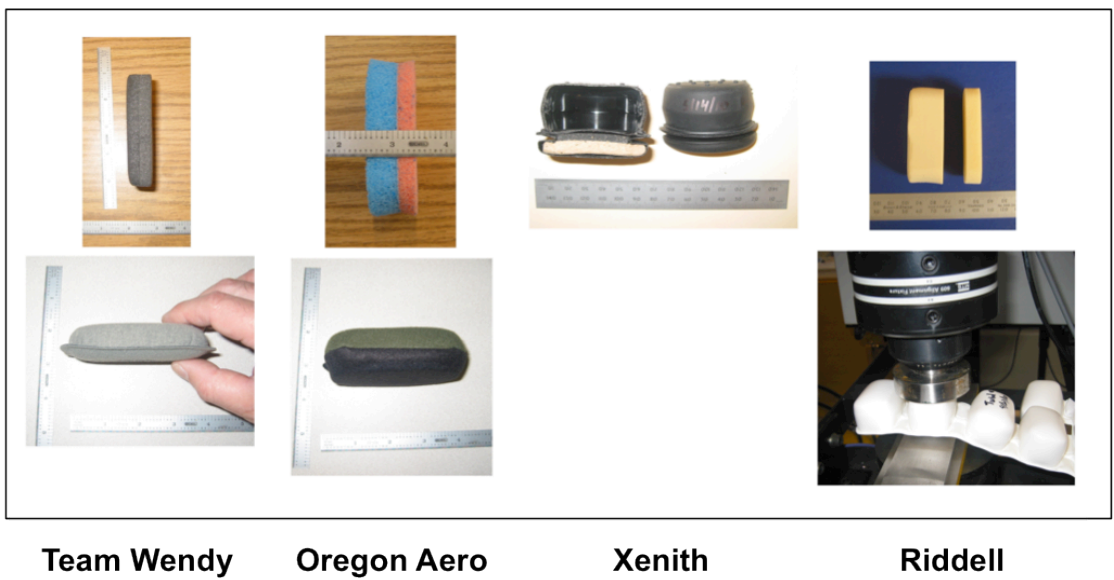

**Team Wendy    Oregon Aero    Xenith    Riddell**

**Figure 1 – Different pads considered in pad impact study: Team Wendy; Oregon Aero; Xenith; Riddell**

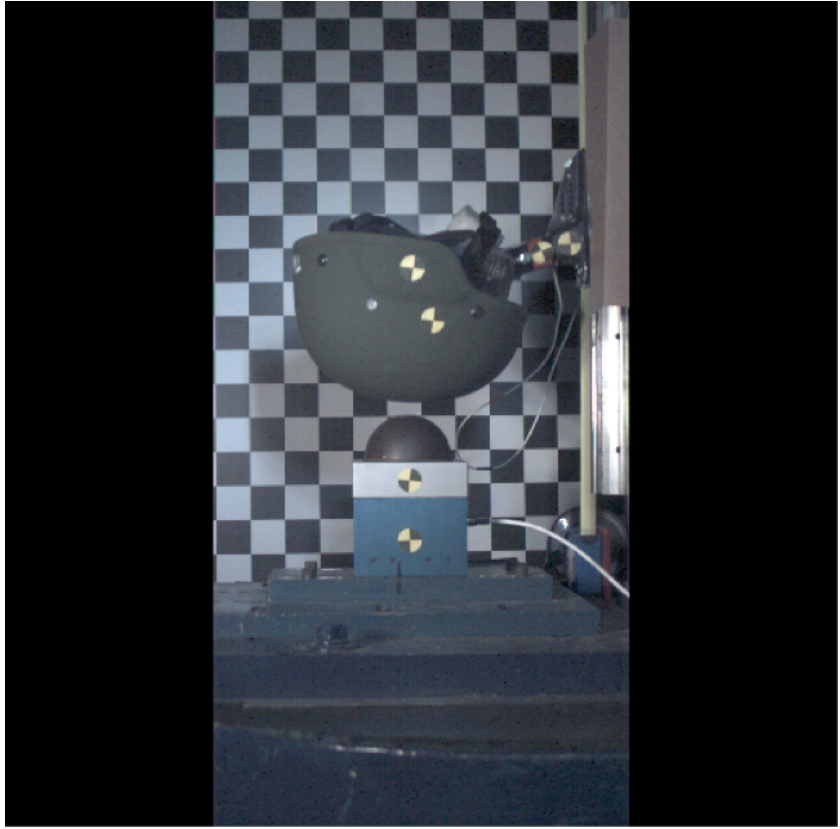

**Figure 2 – An ACH in the USAARL crown impact test just prior to impact on a hemispherical anvil**

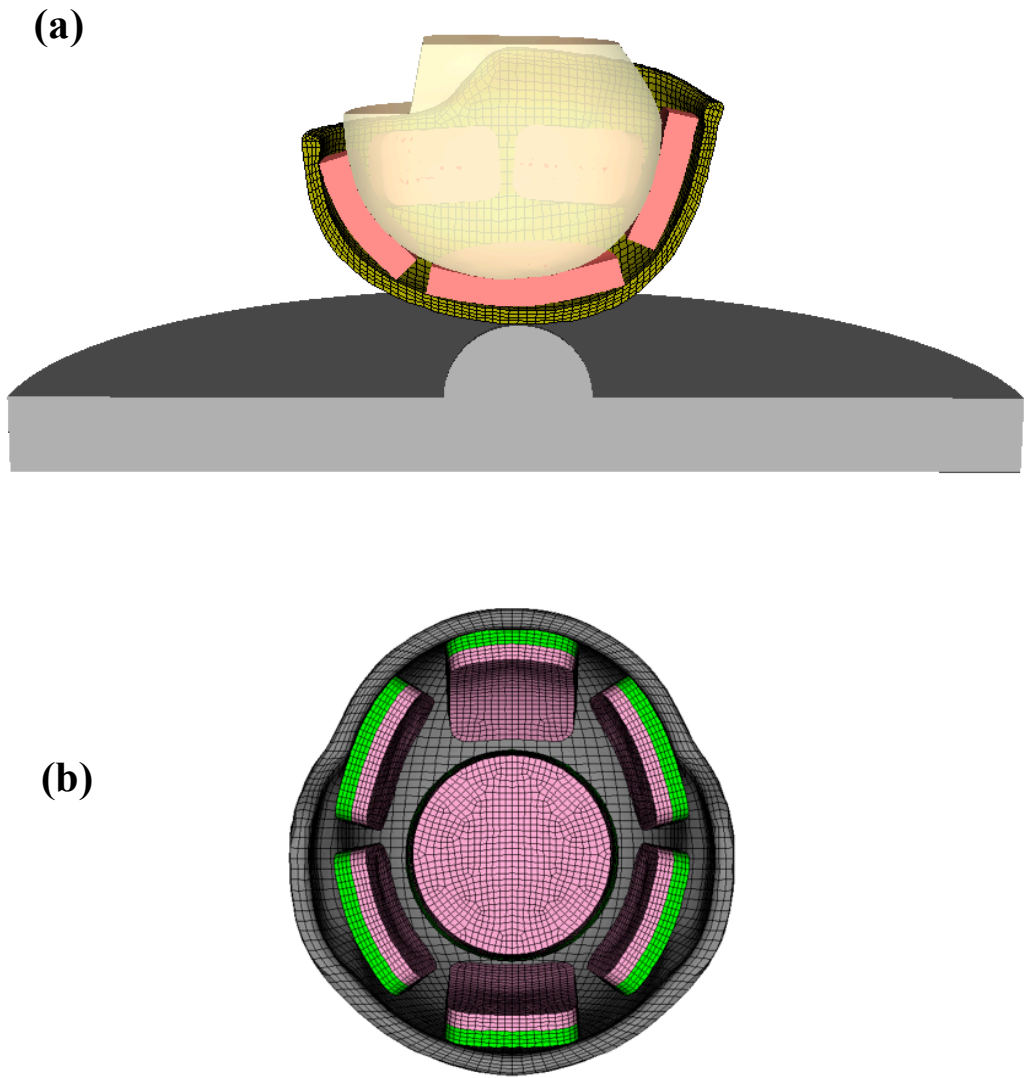

**Figure 3** – Geometry used to simulate the USAARL experiment shown in Figure 2. (a) The half-model that was used for the simulations consists of ~60k elements. Zoning in the helmet is shown. The headform is transparent in order to view the side pads. (b) Interior view of the complete helmet and pads.

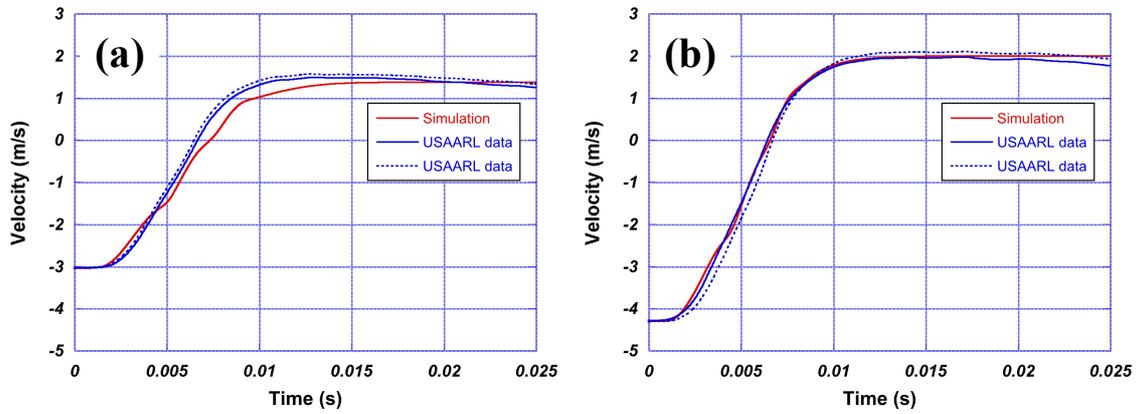

**Figure 4 – Headform center-of-mass velocity in USAARL drop tower experiments and our simulations. (a) 3.05 m/s impact. (b) 4.311 m/s impact.**

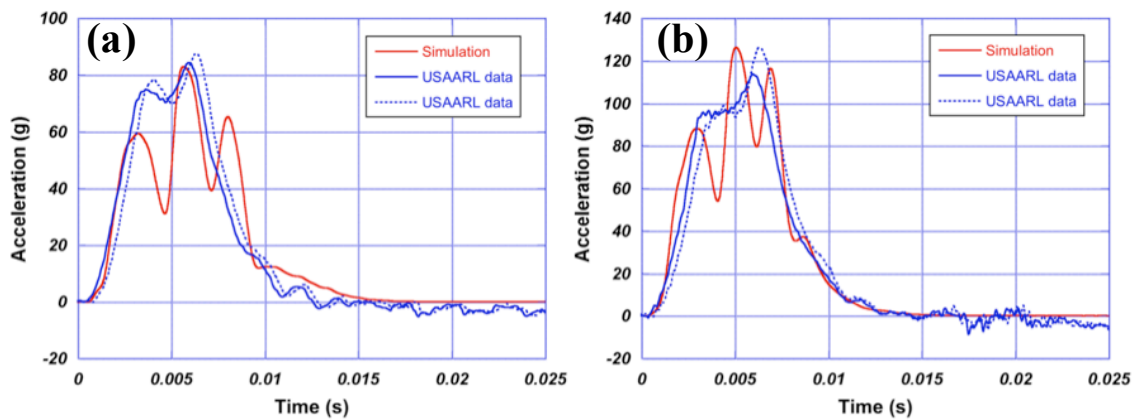

**Figure 5 – Headform center-of-mass acceleration in USAARL drop tower experiments and our simulations. Simulated and experimental peak accelerations and acceleration pulse durations are consistent. (a) 3.05 m/s impact. (b) 4.311 m/s impact.**

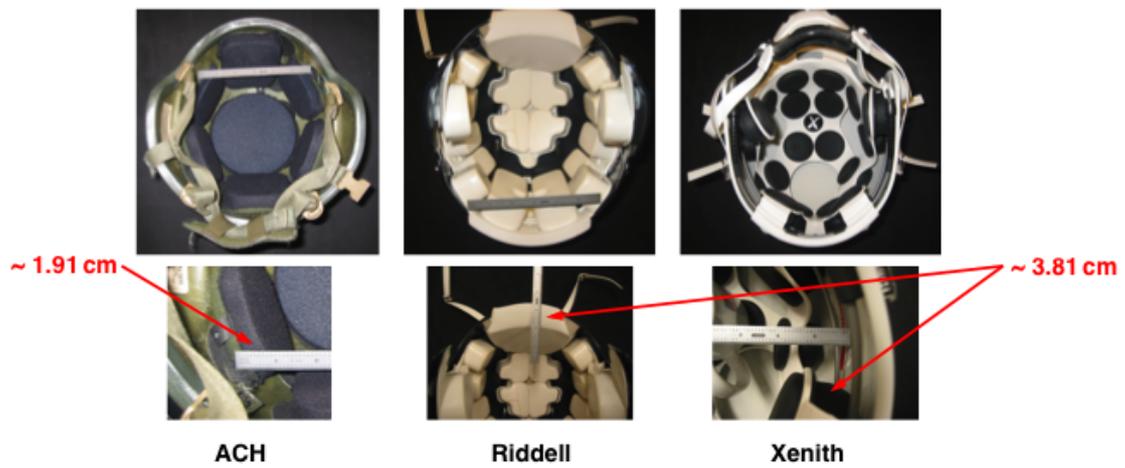

**Figure 6 – Differences in helmet pad size, shape, and distribution between Army and football systems**

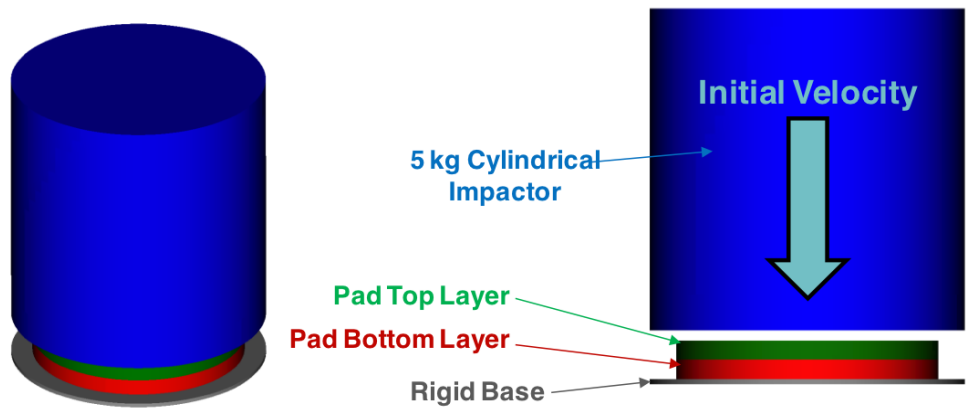

**Figure 7 –Cylinder impact geometry**

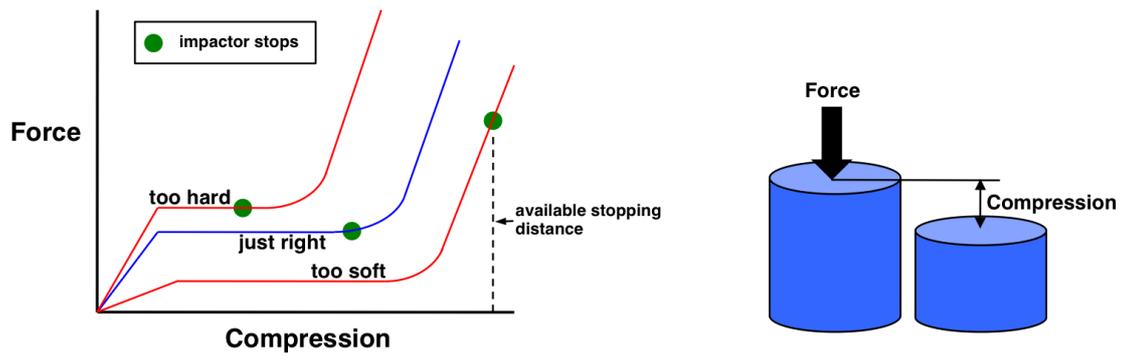

**Figure 8 – General response of foam to compression during an impact. It is assumed that the integral under each curve to the point where the impactor stops (green dot) is equivalent to the initial kinetic energy of the impactor.**

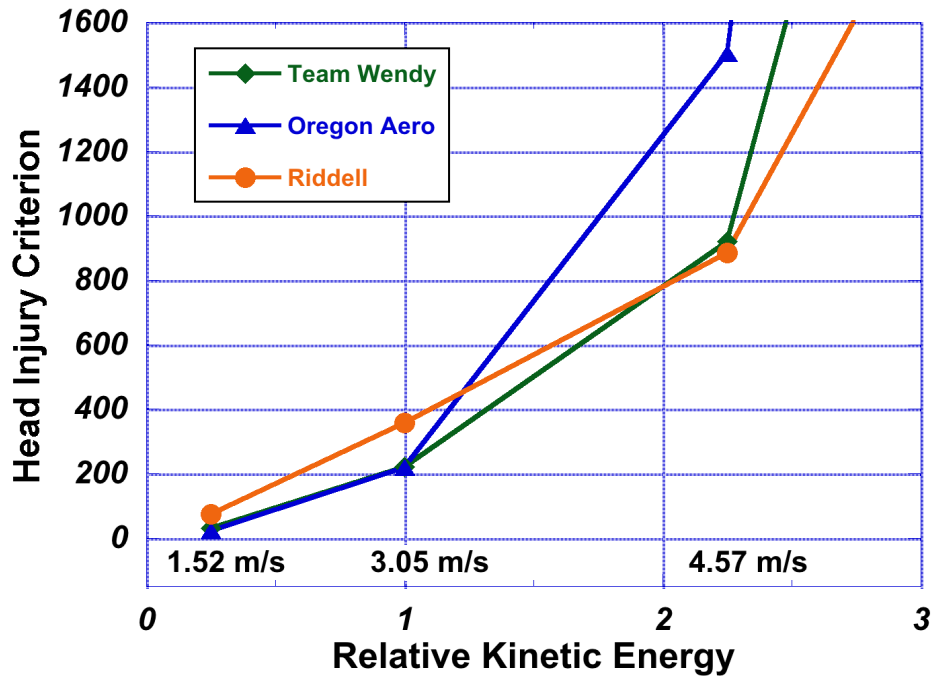

**Figure 9 – HIC as a function of relative kinetic energy for cylinder impact simulations of three different pad systems with identical geometries (1.91 cm thick)**

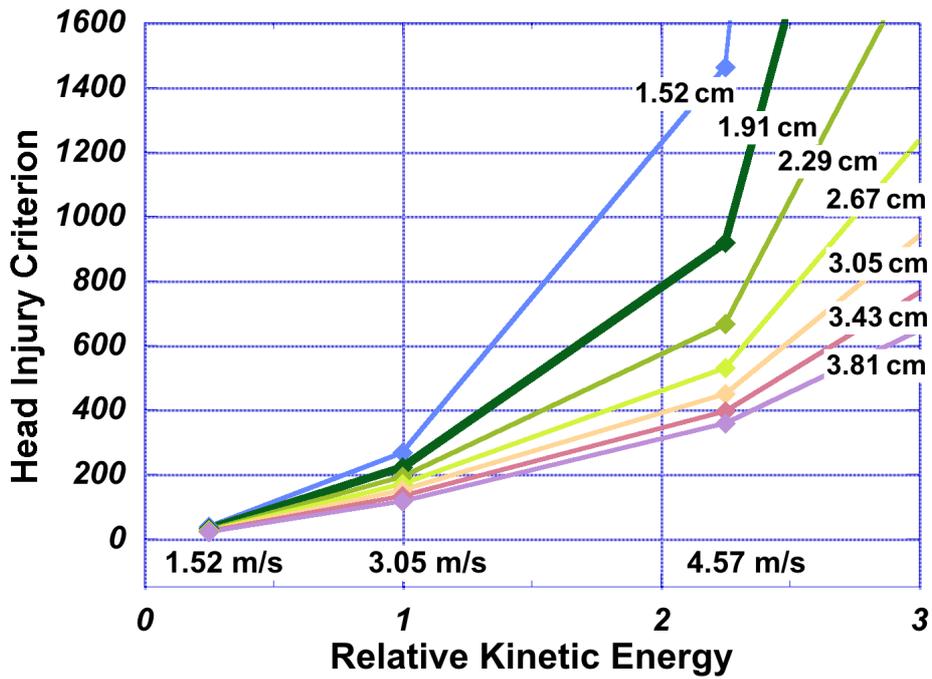

**Figure 10** – Effects of pad thickness for Team Wendy pads in cylinder impact simulations. Increasing the standard 1.91 cm thick pad by 0.38 cm has a significant effect on impact mitigation.

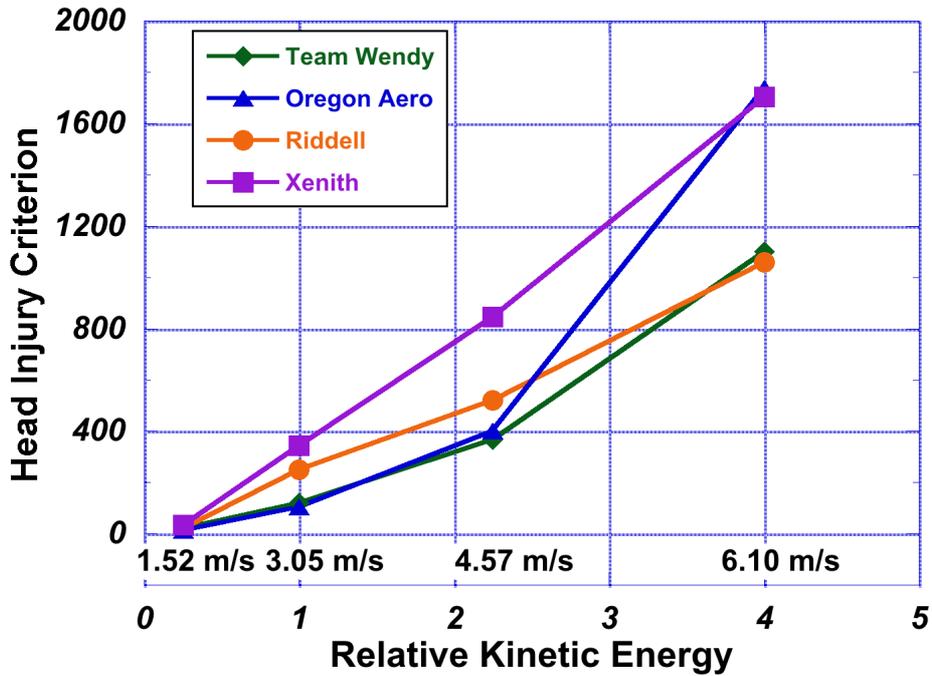

**Figure 11** – HIC as a function of relative kinetic energy for cylinder impact simulations of four different pad systems with identical geometries (3.68 cm thick)

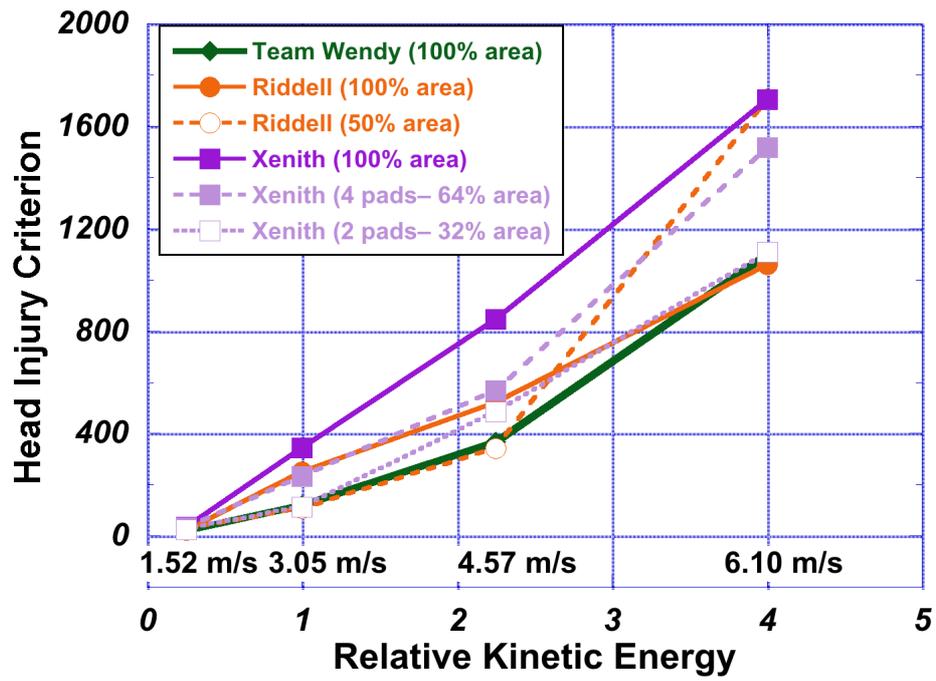

**Figure 12 – Effects of pad area in cylinder impact tests, for a 3.68 cm pad thickenss**